\documentclass[notitlepage,amssymb,amsmath,aps,floatfix,prd,nofootinbib,superscriptaddress,showpacs,preprintnumbers]{revtex4-1}

\usepackage[utf8]{inputenc}

\usepackage{natbib}
\usepackage{graphicx}
\usepackage{epsfig}
\usepackage{amsmath}
\input{epsf}
\usepackage{psfrag}

\usepackage[usenames,dvipsnames]{color}

\newcommand{\beq}{\begin{eqnarray}}
\newcommand{\eeq}{\end{eqnarray}}

\newcommand{\nn}{\nonumber \\}

\usepackage{amsmath,color}
\usepackage{amssymb}
\usepackage{bm}
\usepackage{graphicx}
\usepackage{multirow}

\usepackage{braket}
\usepackage{tikz}
\usetikzlibrary{quantikz2}

\begin{document}

\title{Maximally entangled gluons for any $x$
}

\author{Yoshitaka Hatta}
\email{yhatta@bnl.gov}
\affiliation{Physics Department, Brookhaven National Laboratory, Upton, NY 11973, USA}
\affiliation{RIKEN BNL Research Center, Brookhaven National Laboratory, Upton, NY 11973, USA}

\author{Jake Montgomery}
\email{jake,montgomery@stonybrook.edu}
\affiliation{Department of Physics and Astronomy,
Stony Brook University, New York 11794, USA}

\begin{abstract}
Individual quarks and gluons at small-$x$ inside an unpolarized hadron can be regarded as Bell states in which qubits in the spin and orbital angular momentum spaces are  maximally entangled. Using the machinery of quantum information science, we generalize this observation to all values $0<x<1$ and describe  gluons (but not quarks) as maximally entangled states between a qubit and  a qudit. We introduce the conditional probability distribution $P(l^z|s^z)$ of a gluon's orbital angular momentum $l^z$ given its helicity $s^z$. Restricting to the three states $l^z=0,\pm 1$, which constitute a qutrit, we explicitly compute $P$ as a function of $x$.

\end{abstract}

\maketitle

\section{Introduction}

Recently there have been strong theoretical activities in the application of quantum information science (QIS) to QCD    \cite{Bauer:2022hpo,Abir:2023fpo,DiMeglio:2023nsa}.  Parton physics, which has been traditionally a subject of  perturbative QCD, can potentially benefit from these developments  especially in the context of nucleon tomography and multi-dimensional parton distributions      \cite{Kutak:2011rb,Peschanski:2012cw,Kovner:2015hga,Kharzeev:2017qzs,Hagiwara:2017uaz,Neill:2018uqw,Armesto:2019mna,Ramos:2020kaj,Dvali:2021ooc,Kou:2022dkw,Liu:2022hto,Dumitru:2023qee,Dumitru:2023fih,Hentschinski:2024gaa}. Contrary to the naive picture, partons (quarks and gluons) that constitute a high energy hadron are correlated with one another in position and momentum spaces and in other quantum numbers since they are generated by the successive branchings of higher energy partons with specific probabilities.    
In some cases, these correlations reflect  true quantum mechanical entanglement and can be quantified by various forms of entanglement entropy.  One can then expect that the language and tools of QIS will be  useful for characterizing such phenomena and possibly obtaining novel insights into the nature of partons as a quantum many-body system.

In a recent paper \cite{Bhattacharya:2024sno}, it has been pointed out that partons carrying a small energy fraction $x\ll 1$ of the parent hadron  are maximally entangled. This is not the usual bipartite entanglement that involves two partons, or one between some spatial volume and its `environment'. Instead, it is an entanglement between two internal subspaces of  a {\it single} parton. Namely, a qubit representing the two helicity states ($s^z=\pm 1$ for a gluon and $s^z=\pm \frac{1}{2}$ for a quark in units of $\hbar$) is entangled with another qubit representing the two orbital angular momentum (OAM) states $l^z=\pm 1$. This spin-orbit entanglement is a necessary consequence of the peculiar features of QCD at small-$x$. First,  the possible values of partons'  OAM are effectively restricted to $l^z=\pm 1$  at small-$x$, although in principle it can take any integer values $l^z=0,\pm 1, \pm 2\cdots$. Second, the helicity and OAM of each parton are perfectly {\it anti}-aligned even in an unpolarized or spinless hadron. This results in the formation of the so-called Bell states 
 \beq
 |\Phi_\pm\rangle =\frac{1}{\sqrt{2}}\Bigl(|s^z=1\rangle |l^z=-1\rangle \pm |s^z=-1\rangle |l^z=1\rangle\Bigr).
 \eeq
 ($s^z=\pm \frac{1}{2}$ for quarks.) 
 In the field of quantum optics, similar entangled states of photons have been artificially realized in an ingenious experimental setup \cite{stav}. In QCD, the Bell states are default states into which soft partons are produced. 

In this paper, we discuss whether  similar entangled states exist for partons with generic values $x$ and examine the potential application of QIS in constructing such states.  This is highly nontrivial because neither of the above two features  survives away from the regime $x\ll 1$. Nevertheless, the correlation between partons' helicity and OAM exists for all values of $x$, and this suggests that some degree of entanglement is always present.  We shall show that gluons (but not quarks) are  maximally entangled for all values of $0<x<1$ in a larger Hilbert space of OAM in which more values of $l^z$ are allowed. This can be understood as an entanglement between a qubit and  a {\it qudit} \cite{Gottesman:1998se}---the  multi-dimensional generalization of a qubit. By restricting to the {\it qutrit} case consisting of three states $l^z=0,\pm 1$, we explicitly construct the maximally entangled state as a function of $x$.\footnote{ In the context of high energy and nuclear physics, qutrits have been introduced to simulate the three spin states of spin-1 gauge bosons and vector mesons   \cite{Fabbrichesi:2023cev,Fabbrichesi:2023idl}, three-flavor neutrinos \cite{Siwach:2022xhx,Nguyen:2022snr,Turro:2024shh} and the three isospin states of pions \cite{Beane:2021zvo}. }

\section{Setup}

Consider a quark or a gluon, collectively called a parton, carrying energy  fraction $x$ of a fast-moving hadron in the $z$-direction. The hadron is assumed to be  either unpolarized or spinless. For definiteness, we shall have in mind an unpolarized proton.   Then the quark's two spin states  $s^z=\pm\frac{1}{2}$ are realized with   equal (50\%) probabilities, and similarly for the gluon with $s^z=\pm 1$.   On the other hand, their orbital angular momentum (OAM) can in principle take any integer value $l^z=0,\pm 1, \pm 2, \cdots$. 
Different values of $l^z$ will be realized with different probabilities. If the spin and OAM of individual partons are correlated, the distribution of $l^z$  naturally depends on the value of $s^z$. Namely, one can introduce the conditional probability distribution $P(l^z|s^z)$ normalized as 
\beq
\sum_{s^z}\sum_{l^z=-\infty}^\infty P(l^z|s^z)=1.
\eeq 
$P$ is separately defined for quarks ($q$) and gluons ($g$) and is a function of $x$. We often suppress  these dependencies to avoid cluttering the  indices. In the same vein, we will often  write $s^z,l^z\to s,l$ in equations.  
In terms of the ket vector notation describing  the spin and spatial parts of the parton's  wavefunction  
\beq
|\Phi\rangle&=& 
\sum_{s l}c_{s l}|s\rangle |l
\rangle 
\nn
&=&|+\rangle \Bigl\{a_1|1\rangle+a_0|0\rangle+a_{-1}|-1\rangle + \cdots \Bigr\} +|-\rangle\Bigl\{b_1|1\rangle+b_0|0\rangle+b_{-1}|-1\rangle + \cdots \Bigr\}, \label{ent}
\eeq
we have 
\beq
P(l|+)=|c_{+l}|^2=|a_l|^2, \qquad P(l|-)=|c_{-l}|^2=|b_l|^2.
\eeq
$|\pm\rangle$ denotes  either  quark or gluon spin eigenstates with $s^z=\pm \frac{1}{2}$ or $s^z=\pm 1$, respectively. 
From the normalization condition,  
\beq
\langle \Phi|\Phi\rangle =\sum_{sl}|c_{sl}|^2=\sum_l(|a_l|^2+|b_l|^2)=1. \label{norm}
\eeq
Since the parent hadron is unpolarized, we must require  
\beq
\langle \Phi|s^z|\Phi\rangle=\langle \Phi|l^z|\Phi\rangle=0 .\label{ex}
\eeq
However, the correlation between $s^z$ and $l^z$ is in general nonvanishing 
\beq
\langle \Phi|s^zl^z|\Phi\rangle\equiv \begin{cases} \frac{1}{2} f_{q}(x), \qquad ({\rm quark}) \\ f_g(x). \qquad ({\rm gluon}) \end{cases} \label{ff}
\eeq
The distributions $f_{q,g}(x)$ are proportional to the quark and gluon spin-orbit correlations $C_{q,g}(x)$ \cite{Lorce:2011kd,Hatta:2024otc} which can be rigorously defined in QCD.  We shall specify them later. For the moment let us assume that they are known functions. 
Substituting (\ref{ent}) we find 
\beq
&&\sum_l(|a_l|^2-|b_l|^2)=0, \label{c1}\\
&&\sum_l l(|a_l|^2+|b_l|^2)=0, \label{c2}\\
&&\sum_l l(|a_l|^2-|b_l|^2)=f_{q,g}(x) .\label{c3}
\eeq 
Obviously, there are infinitely many solutions to the system of equations (\ref{c1})-(\ref{c3}). However, 
it has been observed recently  \cite{Bhattacharya:2024sno} that
\beq
  f_{q,g}(x)\approx -1,  \qquad (x
\ll 1) \label{-1}
\eeq
meaning that the spin and OAM of individual partons are perfectly anti-aligned. Moreover,  almost all quarks and gluons at small-$x$ have either $l^z= 1$ or $l^z=-1$. It then follows that (\ref{ent}) reduces to the `Bell states' 
\beq
|\Phi\rangle \approx |\Phi_\pm\rangle=\frac{1}{\sqrt{2}}\Bigl(|+\rangle|\!-\!1\rangle\pm  |-\rangle|1\rangle\Bigr). \label{bell}
\eeq
Equivalently, 
\beq
|a_{-1}|=|b_1|\approx \frac{1}{\sqrt{2}}, \qquad (x\ll 1)
\eeq
for both quarks and gluons, and all the other coefficients are suppressed at small-$x$. 

\section{Gluon as an entangled qubit-qutrit system}

The purpose of this paper is to obtain insights into the nature of the state (\ref{ent}) away from the limit $x\to 0$ where one can no longer argue that only the values $l^z=\pm 1$ need to be considered. Ideally, one would like to determine the coefficients $a_l,b_l$ for all $l$ as a function of $x$. At first sight, this appears to be a formidable  problem that cannot be answered without a detailed understanding of the parton dynamics in QCD. However, in the case of gluons, the problem somewhat simplifies because one  can eliminate $b_l$'s by invoking $PT$ symmetry, the combined operation of parity $P$ and  time-reversal $T$. Under $PT$, the momentum of partons (and the parent hadron)  remains the same,  whereas angular momentum flips signs $s^z,l^z\to -s^z,-l^z$. Since  the two directions of spin and OAM should be completely equivalent in an unpolarized proton, we can  postulate 
\beq
PT|\Phi\rangle =e^{i\varphi}|\Phi\rangle,\label{ttrans}
\eeq
where $e^{i\varphi}$ is an unobservable phase. 
In the basis $PT|+\rangle=|-\rangle$ and $PT|l\rangle =(-1)^l|-l\rangle$,\footnote{This is a choice motivated by the transformation of the typical OAM wavefunction $e^{il\phi}$ where $\phi$ is the azimuthal angle (cf., the Laguerre-Gaussian beam). Under $PT$, $e^{il\phi}\to (e^{il(\phi+\pi)})^* =(-1)^l e^{-il\phi}$. If one chooses a different phase convention, the following equations need to be modified accordingly. In the end the physical result will be the same. } this leads to the relation 
\beq
b_l=e^{-i\varphi}(-1)^l a^*_{-l}.
 \label{tran}
\eeq
Then (\ref{c1}) and (\ref{c2}) are automatically satisfied and (\ref{norm}) and (\ref{c3}) reduce to
\beq
\sum_l|a_l|^2=\frac{1}{2} ,\qquad 
\sum_l l|a_l|^2=\frac{1}{2}f_{g}(x). \label{set}
\eeq
Note that, for a single quark  which is a fermion, the so-called Kramers degeneracy  associated with time-reversal symmetry forbids one from writing down a relation analogous to (\ref{ttrans}). In the following, we focus on the gluon case, leaving the quark case to future work. 

Even after this simplification, (\ref{set}) still appears intractable since an infinite sum over $l^z$ is involved. 
To make progress, we simplify the problem by assuming that the Hilbert space of OAM is limited to the states $l^z=0,
\pm 1$. This may sound a drastic approximation, but actually, it is  physically motivated. First, it is counter-intuitive to find many partons with $|l^z|
\ge 2$ inside a proton with spin $\frac{1}{2}$. Moreover, if one  estimates  OAM  $|l^z|\sim |\vec{k}\times \vec{b}|$ classically  using $|\vec{k}|
\lesssim \Lambda_{QCD}\sim 200$ MeV and $|\vec{b}|\lesssim 1$ fm (the size of a proton), then indeed $|l^z|
\lesssim 1$. Under this assumption, (\ref{set}) reduces to 
\beq
|a_1|^2+|a_0|^2+|a_{-1}|^2 = \frac{1}{2} , \qquad 
|a_1|^2-|a_{-1}|^2 =\frac{ f_{g}(x)}{2} .\label{ex2}
\eeq
Since there are two equations for three unknowns, the solution of (\ref{ex2}) is still not unique. We however show that, given $f_{g}(x)$, the complex vector $(a_1,a_0,a_{-1})$ can be  determined up to a phase.

For this purpose, it is convenient to switch to the language of quantum information science. The Bell state (\ref{bell}) is  a maximally entangled state of two qubits, one in the spin space $|\pm\rangle$ and the other in the OAM space $|\!\pm 1\rangle$.  The latter interpretation is possible only when $x\ll 1$ where  only the states $l^z=\pm 1$ matter. If we  include   $d>2$  values of $l^z$, we need to introduce a `qudit' \cite{Gottesman:1998se} whose Hilbert space is spanned by $d$-dimensional complex vectors. 
One may also consider a `qumode' \cite{Braunstein:2005zz} in the limit $d\to \infty$ since $l^z$ can in principle take any integer values. 
In the present approximation, $d=3$, and the triplet of states $l^z=0,\pm 1$ becomes a `qutrit'.  
Therefore, we shall regard a gluon with $0<x<1$ as an entangled system between a qubit and a qutrit.

Let us discuss the entanglement entropy of this qubit-qutrit system. The reduced density matrix traced over the OAM space  is 
\beq
{\rm Tr}_l|\Phi\rangle\langle \Phi| =\sum_l \langle l|\Phi\rangle\langle \Phi|l\rangle = \begin{pmatrix} \sum_l|a_l|^2 & \sum_l a_l b_l^* \\ \sum_l a_l^*b_l & \sum_l |b_l|^2\end{pmatrix} = \begin{pmatrix} \frac{1}{2}&  e^{i\varphi}\sum_l (-1)^l a_l a_{-l} \\ e^{-i\varphi}\sum_l (-1)^l a_l^*a_{-l}^* & \frac{1}{2}\end{pmatrix} .
\eeq
This can be diagonalized with the eigenvalues 
\beq
\lambda_\pm = \frac{1}{2} \pm \left|\sum_l (-1)^la_la_{-l}\right|.
\eeq
The entanglement entropy  
\beq
S=-\sum_\pm \lambda_\pm \ln \lambda_\pm, 
\eeq
 takes a maximal value  $S=\ln 2\approx 0.69$ when 
\beq
\sum_l (-1)^la_la_{-l}=a_0^2+2\sum_{l=1}^\infty (-1)^la_la_{-l}=0 .\label{max}
\eeq
The Bell states (\ref{bell}) are  maximally entangled  because $a_{l}=0$ for $l^z\neq -1$.

Our crucial observation is that the states at $0<x<1$ are also maximally entangled with $S=\ln 2$. The argument goes as follows. 
We regard the probabilities $|a_l|^2$ as  averages over gluons with the same value of $x$. 
We can then expect that the complex vector $(a_1,a_0,a_{-1})$ varies smoothly as a function of $x$.   Since the norm of the vector  is conserved (\ref{ex2}), this variation can be implemented by  $x$-dependent unitary transformations 
 \beq
 a_{l}\to U_{ll'}a_{l'}  ,
\eeq
 where $U$ is a $x$-dependent 3$\times$3 unitary matrix  $UU^\dagger=U^\dagger U=1$ continuously connected to the unit matrix. Note that, in general, the quantum evolution of the qubit-qutrit system is represented by six-dimensional unitary matrices $W$ acting on the six-dimensional complex vector $\{c_{sl}\}=(a_1,a_0,a_{-1},b_1,b_0,b_{-1})$. However, in the present problem $a_l$ and $b_l$ are related
 due to $PT$ symmetry. This means that $W$ factorizes into a tensor product $W=V\otimes U$ where $V$ is a 2$\times$2 matrix acting on the spin space (qubit). Namely, $\{c_{sl}\}$ transforms as 
 \beq
 c_{sl}\to V_{ss'}U_{ll'}c_{s'l'}.
 \eeq
 Consistency with (\ref{tran}) requires  
 
\beq
&&V_{hh'}^* =V_{-h,-h'}, \label{v}\\
&& U^*_{ll'}=(-1)^{l+l'}U_{-l,-l'}.\label{con}
\eeq
The only 2$\times$2 unitary matrix that satisfies  
 (\ref{v}) and is continuously connected to the unit matrix  takes the form

\beq
V=\begin{pmatrix} e^{i\theta} & 0 \\ 0 & e^{-i\theta} \end{pmatrix}.
\label{vmat}
\eeq 
Physically, this represents the rotation of the gluon's linear polarization vector around the $z$-axis $\text{R}_\text{z}(\theta) = \exp{\left[i \theta \sigma_z\right]}$.
On the other hand, a 3$\times$3 unitary matrix $U$ can satisfy  (\ref{con}) more nontrivially. 

Importantly, $V,U$ are {\it local} unitary transformations that act separately on the qubit and qutrit subsystems. 
It is  known that local unitary transformations preserve the entanglement entropy \cite{Nielsen:2012yss}. Moreover, the unitary evolution is reversible. 
Since  the Bell state (\ref{bell}) in the limit $x\to 0$ is maximally entangled, states with $0<x<1$ reached from the Bell state via successive local unitary transformations are also maximally entangled. 
Note that if $c_{sl}$ evolves in the most general way via  six-dimensional unitary matrices $W$, the evolution modifies the entanglement entropy  between the two subsystems. Presumably this is what happens in the quark sector.

The above observation is sufficient to determine the vector $(a_1,a_0,a_{-1})$. It follows from (\ref{ex2}) that 
\beq
2|a_1|^2+|a_0|^2=\frac{1+f_{g}(x)}{2}\equiv \epsilon. 
\eeq
We parameterize the solution as 

\beq
a_1= e^{i\chi'}\sqrt{\frac{1-\kappa}{2}\epsilon}, \quad a_0=e^{i\chi}\sqrt{\kappa \epsilon}, \qquad a_{-1}=\sqrt{\frac{1}{2}-\frac{1+\kappa}{2}\epsilon},
\eeq
where $0<\kappa<1$. 
To maximize the entropy $S$, we require 
\beq
0 = a_0^2-2a_1a_{-1} = e^{2i\chi}\kappa\epsilon -2e^{i\chi'}\sqrt{ \frac{1-\kappa}{2}\epsilon \left(\frac{1}{2}-\frac{1+\kappa}{2}\epsilon\right)}
\eeq
This is satisfied if $\chi'=2\chi$ and 
\beq
\kappa=  1-\epsilon,
\eeq
so that 
\beq
a_1=e^{2i\chi}\frac{\epsilon}{\sqrt{2}}, \quad a_0=e^{i\chi}\sqrt{(1-\epsilon)\epsilon}, \quad a_{-1}= 
\frac{1-\epsilon}{\sqrt{2}}. \label{ave}
\eeq
We thus obtain 
\beq
&&P(1|+)=P(-1|-)=\frac{\epsilon^2}{2}, \nn &&P(0|+)=P(0|-)=(1-\epsilon)\epsilon, \\ && P(-1|+)=P(1|-)=\frac{(1-\epsilon)^2}{2}. \nonumber
\eeq
Moreover, from (\ref{con}) and (\ref{ave}) 
 we can determine the unitary matrix $U$ up to a phase 
\beq
\begin{pmatrix} a_1 \\ 
a_0 \\ a_{-1}
\end{pmatrix}
&=&e^{i\theta}\begin{pmatrix} 1-\epsilon & e^{i\chi} \sqrt{2\epsilon(1-\epsilon)} & e^{2i\chi}\epsilon \\ -e^{-i\chi} \sqrt{2\epsilon(1-\epsilon)} & 1-2\epsilon & e^{i\chi} \sqrt{2\epsilon(1-\epsilon)} \\ 
e^{-2i\chi}\epsilon & -e^{-i\chi} \sqrt{2\epsilon(1-\epsilon)} & 1-\epsilon 
\end{pmatrix} \begin{pmatrix} 0 \\ 0 \\ 
\frac{1}{\sqrt{2}}\end{pmatrix}.  \label{fe}
\eeq
The Bell states (\ref{bell}) are recovered in the limit $\epsilon\to 0$, while the limit $\epsilon \to 1$ corresponds to  the other two Bell states
\beq
|\Phi\rangle \to |\Psi_{\pm}\rangle=\frac{1}{\sqrt{2}}\Bigl( |+\rangle|1\rangle \pm |-\rangle|-1\rangle \Bigr). \label{other}
\eeq
In general, when $0<\epsilon<1$, we have a  maximally entangled state between a qubit and a qutrit.  
Introducing the `time' variable $t$ 
\beq
1-2\epsilon =-f_{g}(x) \equiv \cos t,
\label{timedef}
\eeq
we can write   
\beq U= \begin{pmatrix} \frac{1+\cos t}{2} & e^{i\chi} \frac{\sin t}{\sqrt{2}} & e^{2i\chi}\frac{1-\cos t}{2}\\ -e^{-i\chi} \frac{\sin t}{\sqrt{2}} & \cos t & e^{i\chi} \frac{\sin t}{\sqrt{2}} \\ 
e^{-2i\chi}\frac{1-\cos t}{2} & -e^{-i\chi} \frac{\sin  t}{\sqrt{2}}  & \frac{1+\cos t}{2}
\end{pmatrix} \equiv  e^{-iHt} ,
\eeq
where  
\beq
H&=& \frac{i}{\sqrt{2}}\begin{pmatrix} 0 & e^{i\chi} & 0 \\ -e^{-i\chi} & 0 & e^{i\chi} \\ 0 & -e^{-i\chi} & 0\end{pmatrix} \nn 
&=& \frac{i}{\sqrt{2}}\Bigl\{ e^{i\chi} |1\rangle\langle 0|-e^{-i\chi} |\!-\!1\rangle\langle 0|-e^{-i\chi} |0\rangle\langle 1| +e^{i\chi} |0\rangle\langle -1| \Bigr\} \label{hami}
\eeq
is the Hamiltonian which acts on the qutrit and generates the unitary evolution. It is worth noting that, for a choice of $\chi = 0$, the Hamiltonian is a generator of rotations, $-L_y$, and the evolution therefore is a rotation $\text{R}_\text{y}(-t)$.

\begin{figure}[t]
\begin{center}
\includegraphics[width=0.7\textwidth]{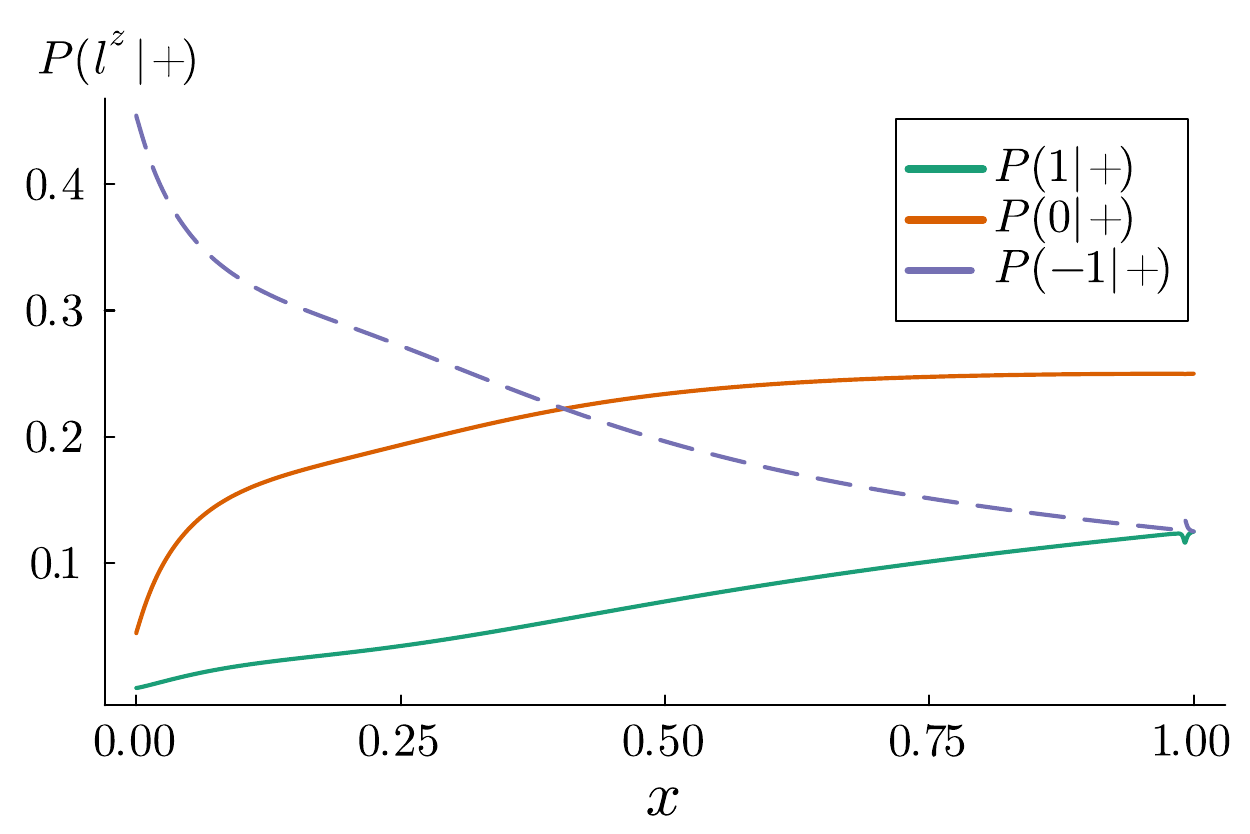}
\end{center} 
\caption[*]{Conditional probability distributions for the truncated number of orbital angular momentum states given the spin state is $\ket{+}$. The sum of the three curves is normalized to $\frac{1}{2}$. }
\label{2}
\end{figure}

\section{Numerical results}

We finally come to the functions $f_{q,g}(x)$. They are defined as the quark and gluon spin-orbit correlations $C_{q,g}(x)$ divided by the number of partons 
\beq
f_q(x) =\frac{2C_q(x)}{q(x)}, \qquad f_g(x) = \frac{C_g(x)}{g(x)}, \label{fqg}
\eeq
 where $q(x)$ and $g(x)$ are the unpolarized quark and gluon parton distribution functions (PDFs). The factor of 2 is conventional, see (\ref{ff}).  
Focusing on the gluon case, we approximate it as \cite{Bhattacharya:2024sck,Hatta:2024otc}
\beq
 C_g(x) \approx -2x\int_x^1 \frac{dz}{z^2}g(z), \label{ww}
\eeq
The complete, exact QCD formula  includes the polarized gluon PDF and the so-called genuine twist-three distributions (the correlation functions of three gluon fields $\langle FFF\rangle$), see  \cite{Hatta:2024otc}.

When $x\ll 1$, the unpolarized PDF shows the Regge behavior 
\beq
g(x)\propto \frac{1}{x^{1+\lambda}},
\eeq
with $\lambda\ll 1$. Inserting this into (\ref{ww}), we find 
\beq
 C_g(x)\approx -\frac{2}{2+\lambda}g(x), \qquad f_{g}(x) \approx -\frac{2}{2+\lambda}\approx -1, \label{fg}
\eeq
as already mentioned in (\ref{-1}). For generic values of $x$,  we use the CT18 NLO gluon PDF \cite{Hou:2019qau} (with the help of \cite{Clark:2016jgm}) at the renormalization scale $Q^2=10$ GeV$^2$  to evaluate $f_{g}(x)$.  
Fig.~\ref{2} shows  
\beq
P(1|+)=\frac{(1+f_{g}(x))^2}{8}, \quad P(0|+)= \frac{1-f_{g}^2(x)}{4}, \quad P(-1|+)
=\frac{(1-f_{g}(x))^2}{8},
\eeq
as a function of $x$. Errors in the PDF have been neglected in this exploratory study. However, we anticipate  that they will partly cancel in the ratio (\ref{fqg}).  In the present approximation (\ref{ww}), $-1<f_{g}(x)<0$, or $0<\epsilon<\frac{1}{2}$. This means that  the other two Bell states (\ref{other}) are not realized in QCD, unless  $f_g(x)$ flips signs  in the large-$x$ region after including the polarized gluon PDF and the genuine twist-three corrections to (\ref{ww}) \cite{Hatta:2024otc}.  

Since $\lambda$ is small but nonzero in QCD, there is a small deviation from the Bell states $P(0|+)\neq 0$ even in the $x\to 0$ limit. Typically, $0.1<\lambda<0.3$, and within this range $\lambda$ is an increasing function of the renormalization scale $Q$ \cite{H1:2001ert}. Thus the deviation also increases with $Q$. 
Turning to the large-$x$ region $0.5<x<1$, we see that gluons tend to have $l^z=0$. This is reasonable because these energetic gluons carry the majority of the proton momentum, which implies that their energy is   primarily used as linear (rather than angular) momentum. As $x\to 1$, $f_g(x)$ vanishes linearly $f_g(x)\propto 1-x$. This is a consequence of the standard parameterization $g(x)\propto (1-x)^\alpha$ with  $\alpha>0$. Accordingly,    $P(1|+)\approx P(-1|+)\approx\frac{1}{8}$ and $P(0|+)\approx\frac{1}{4}$ when $x\approx 1$.

\section{Conclusions}

In this paper we have proposed an unconventional description of   gluons in an unpolarized/spinless hadron as maximally entangled states between a qubit and a qutrit representing the spin and OAM spaces, respectively.   In the $x\to 0$ limit, the qutrit reduces to a qubit, and the Bell states  \cite{Bhattacharya:2024sno} are recovered.  The present work is therefore the generalization of \cite{Bhattacharya:2024sno} to arbitrary values of $x$. Our quantum mechanical description naturally lead to  the conditional probability distributions $P(l^z|s^z)$. Such probabilities may be properly formulated and implemented in realistic wavefunction-based  approaches to hadron structure such as  the light-front approach \cite{Shuryak:2021yif,Xu:2024sjt}.  We have shown that $P$ can be essentially computed using the machinery of quantum information science, and the only dynamical input from QCD  is the gluon spin-orbit correlation $C_g(x)$ which  can be extracted from high energy experiments \cite{Bhattacharya:2018lgm,Boussarie:2018zwg,Bhattacharya:2024sck}. It remains to be seen whether $P$ can be directly connected to experimental observables.

We have found that each gluon has entanglement entropy $\ln 2$ irrespective of the value of $x$. Since different gluons do not interact to first approximation, the density matrix in the space of many gluons is block diagonal. Consequently,  the total entropy of gluons per rapidity $y=\ln \frac{1}{x}$ and in a transverse area $A_\perp$ is  given by 
\beq
\frac{dS(x)}{d\ln \frac{1}{x}}= \ln 2 \frac{A_\perp}{\pi R^2} x\,g(x) , \label{x}
\eeq 
where $R$ is the proton radius. 
In Deep Inelastic Scattering with photon virtuality $Q^2$, we may take $A_\perp \sim 1/Q^2$. It is interesting to notice that 
(\ref{x}) parameterically agrees with the results in  \cite{Kutak:2011rb,Kovner:2015hga,Hagiwara:2017uaz,Dumitru:2023qee} although the definitions of entropy are rather different. At small-$x$, the formula predicts a rapid rise $1/x^\lambda$ of the entropy with decreasing $x$. We however emphasize that (\ref{x}) is not limited to the small-$x$ region.

\begin{figure}
    \centering \includegraphics[width=0.5\linewidth]{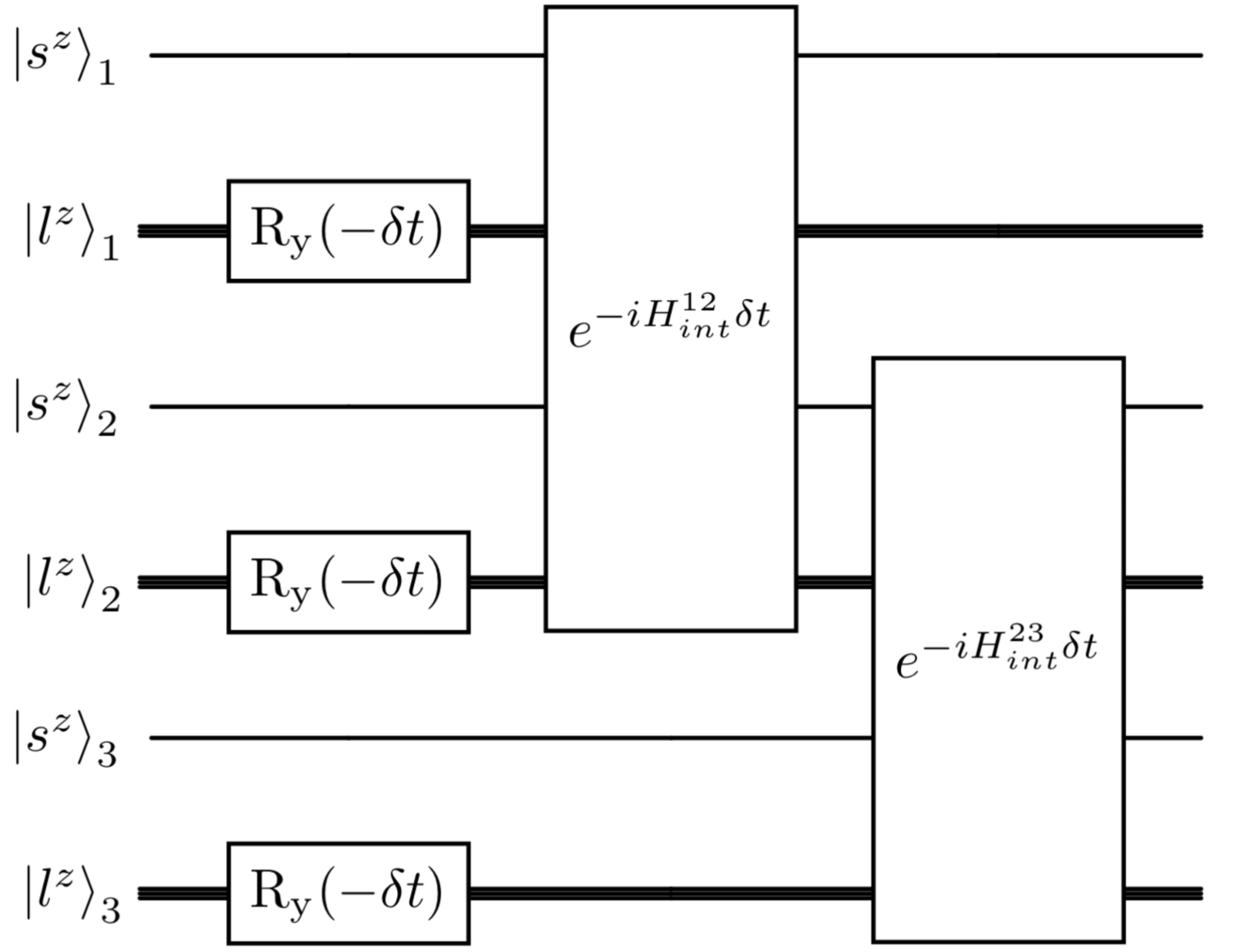}
    \caption{A circuit diagram of a trotterization of evolution operator correspond to the total Hamiltonian (\ref{htot}). Thin lines represent qubits which evolve with (\ref{vmat}). Thick lines represent qutrits which evolve with (\ref{hami}). The phases $\theta,\chi$ are unobservable and set to zero for simplicity, leaving the action of (\ref{vmat}) on the qubits to be trivial. }
    \label{fig:enter-label}
\end{figure}

There are a number of directions for future research: 
\begin{itemize}

\item  The present result is  limited by the assumption $|l^z|\le 1$. Inclusion of higher $|l^z|$ values enlarges the Hilbert space of OAM and poses technical  challenges. For example, when $|l^z|\le 2$, the Hilbert space becomes five-dimensional and the resulting  qubit-{\it qudit} system becomes ten-dimensional. The factorization $W=U\otimes V$ still holds and the entanglement entropy is still maximal, but $U$ is now a 5$\times$5 matrix which is difficult to constrain.  Note that the 3$\times$3 solution is also a solution of the 5$\times$5 problem, but there may be other branches of solutions.  

\item  Another challenge is  the quark sector. That quarks are also in the Bell state (\ref{bell}) in the $x\to 0$ limit appears to be a consequence of the fact that small-$x$ quarks are predominantly generated by the branching of small-$x$ gluons $g\to q\bar{q}$ \cite{Bhattacharya:2024sno}. Quarks are entangled because  the parent  gluons are entangled.  The same reasoning does not hold in the large-$x$ region dominated by the pre-existing valence quarks. Interestingly, in the present quantum mechanical treatment, this difference between quarks and gluons has manifested as one between fermions and bosons. For fermions, Kramers degeneracy does not allow for a simple relation between $a_l$ and $b_l$.   Presumably, the entanglement entropy of  quarks starts out low when $x \sim 1$ and increases toward $\ln 2$ as $x$ is decreased. This remains to be carefully investigated. 

\item
 If the proton is longitudinally polarized,  (\ref{ex}) should be replaced by 
\beq
\langle \Phi|s^z|\Phi\rangle = \frac{\Delta g(x)}{g(x)}, \qquad \langle \Phi|l^z|\Phi\rangle = \frac{L_g(x)}{g(x)},
\eeq
where $\Delta g(x)$ is the gluon helicity PDF and $L_g(x)$ is the gluon OAM distribution \cite{Hatta:2012cs}. For consistency an additional term  should be included  in (\ref{ww})
\beq
C_g(x) \approx -2x\int_x^1 \frac{dz}{z^2}g(z)+x\int_x^1 \frac{dz}{z}\Delta g(z),
\eeq
still neglecting the genuine twist-three terms \cite{Hatta:2024otc}. 
Since the added term is suppressed when $x\ll 1$ \cite{Bhattacharya:2024sck}, we still have the Bell states  in the small-$x$ limit. However,  (\ref{tran}) no longer holds and this suggests that the generic states  $0<x<1$  are not maximally entangled even for gluons. Additional dynamical inputs from QCD may be needed to fully determine the quantum state.

\item  Finally, in this paper we have considered the quantum state of individual gluons, neglecting possible correlations between the spin and OAM of different gluons as encoded, for example, in double parton distributions. In a quantum mechanical treatment, such higher order correlations may be described by a residual interaction Hamiltonian $H_{int}$ acting on pairs of gluons. One can then consider the evolution of a system of many gluons with the total Hamiltonian 
\beq
H_{tot}=\sum_i H^i + \sum_{ij}H^{ij}_{int},
\label{htot}
\eeq
where $H^i$ is the one-body Hamiltonian consisting of (\ref{vmat}) and (\ref{hami}) for the $i$-th gluon.  

Whether or not this evolution 
can be treated 
analytically is an open question, but should the need for simulation arise, 
classical computing methods will eventually struggle to be a viable method. 
Quantum computing is well suited to treat unitary evolution in quantum systems since the Hilbert space 
accessible scales exponentially with the amount of resources, and gates are naturally
unitary operators. Specifically, hybrid quantum computers hold promise for directly simulating the spin-OAM Hilbert spaces using qubits and qutrits/qudits/qumodes. Indeed, the problem (\ref{htot}) is similar to
other systems that map the dynamical variables to qubit-qudit/qumode entangled states, such as those recently studied in  \cite{Araz:2024kkg} \cite{Zache:2023cfj} \cite{Illa:2024kmf} on  hybrid quantum computers. Motivated by this analogy, we draw in Fig.~\ref{fig:enter-label}  a schematic circuit diagram for three gluons with  unspecified inter-gluon gates representing the interaction $H_{int}$. However, this is an oversimplified picture especially because in reality the number of gluons is not a constant in `time'  (see (\ref{timedef})). It remains to be seen whether the further developments in QIS will prove beneficial for QCD parton physics.  

\end{itemize}

\section*{Acknowledgements}
J. M. thanks the EIC theory institute at Brookhaven National Laboratory, where this work was initiated,  for hospitality.   
Y.~H. was supported by the U.S. Department of Energy under Contract No. DE-SC0012704, and also by  Laboratory Directed Research and Development (LDRD) funds from Brookhaven Science Associates. J.~M. was supported in part by the DOE, Office of Science, Office
of Nuclear Physics under contract
No. DE-SC0024358.

\bibliography{ref}

\end{document}